\documentstyle[aps,prl,multicol,epsf]{revtex}
\title{\bf The Dynamics of Efficiency: A Simple Model}

\begin{document}
\author{Satya N. Majumdar$^{1,2}$ and P. L. Krapivsky$^3$}
\address{$^1$Laboratoire de
   Physique Quantique (UMR 5626 du CNRS), Universit\'e Paul Sabatier,
   31062 Toulouse Cedex, France}
\address{$^2$Tata Institute of
   Fundamental Research, Homi Bhabha Road, Mumbai-400005, India}
\address{$^3$Center for Polymer Studies and Department of Physics, 
   Boston University, Boston, MA 02215, USA}
\maketitle

\begin{abstract}
  We propose a simple model that describes the dynamics of efficiencies
  of competing agents.  Agents communicate leading to increase of
  efficiencies of underachievers, and an efficiency of each agent can
  increase or decrease irrespectively of other agents.  When the rate of
  deleterious changes exceeds a certain threshold, the economy falls
  into a stagnant phase.  In the opposite situation, the economy
  improves with asymptotically constant rate and the efficiency
  distribution has a finite width.  The leading algebraic corrections to
  the asymptotic growth rate are also computed.
  
\end{abstract}

\begin{multicols}{2}
\noindent  
Non-equilibrium statistical mechanics is being increasingly applied to
diverse fields outside physics, ranging from biology and computer
science to finance and social science\cite{comp,fin,soc}.  Indeed, the
framework of non-equilibrium statistical mechanics is ideally suited for
describing systems composed of many units that interact according to
simple rules and exhibit a complex large-scale behavior.  Thus, the
important task is to construct simple stochastic models incorporating
basic characteristics of the dynamics of systems under study which can
then be analyzed by employing existing tools of non-equilibrium
statistical mechanics.  The hope is that these models can reproduce
essential features of the original systems and can help to formulate
relevant questions and enhance understanding of the dynamics of these
systems.

In this paper, we propose a simple model that mimics the dynamics of
efficiencies of competing agents.  These agents could be airlines,
travel agencies, insurance companies, etc. In today's competing global
economy, the performance of a company is continuously judged in the
market and the index of performance depends on how efficient the company
is.  Rather than trying to incorporate all details of performances of
competing agents, we choose a model that accounts for the dynamics of
efficiency in the simplest form.  We represent the efficiency of each
agent by a single nonnegative number.  The efficiency of every agent
can, independent of other agents, increase or decrease stochastically by
a certain amount which we set equal to unity.  In addition, the agents
interact with each other which is the fundamental driving mechanism for
economy.  We assume that the interaction equates the efficiencies of
underachievers to the efficiencies of better performing agents.

The efficiency model formalizing the above features is defined as
follows.  Let $h_i(t)$ is the efficiency of agent $i$ at time $t$.
Efficiencies $h_i$'s are non-negative integer numbers which evolve
stochastically. Specifically, in an infinitesimal time interval $\Delta
t$, every $h_i(t)$ can change as follows:
   
\begin{itemize} 
\item[(i)] $h_i(t)\to {\rm max}[h_i(t),h_j(t)]$ with probability $\Delta
  t$, where the agent $j$ is chosen randomly. This move is due to the
  fact that each agent tries to equal its efficiency to that of a
  better performing agent in order to stay competitive.
  
\item[(ii)] $h_i(t) \to h_i(t)+1$ with probability $p\Delta t$.  This
  incorporates the fact that each agent can increase its
  efficiency, say due to innovations, irrespective of other agents.
  
\item[(iii)] $h_i(t) \to h_i(t)-1$ with probability $q\theta(h_i(t))
  \Delta t$, where $\theta(x)$ is the Heaviside step function. This
  corresponds to the fact that each agent can loose its efficiency due
  to unforeseen problems such as labour strikes. Note, however, that
  since $h_i(t)\ge 0$, this move can occur only when $h_i(t)\ge 1$.  

\item[(iv)] With probability $1-[1+p+q\theta(h_i(t))]\Delta t$, the
  efficiency $h_i(t)$ remains unchanged.
\end{itemize}  

This efficiency model exhibits rich phenomenology.  In particular, the
system undergoes a delocalization phase transition as the parameters $p$
and $q$ are varied.  There exists a critical line $p_c(q)$ in the
$(p,q)$ plane such that for $p>p_c(q)$, the average efficiency increases
linearly with time, $\langle h \rangle\sim vt$ for large $t$.  We shall
determine exactly the rate $v(p,q)$ of the average efficiency growth.
For $p\le p_c(q)$, the economy is stagnant, i.e., the efficiency
distribution becomes stationary in the large time limit.  This
delocalization (or depinning) phase transition is dynamical in nature
and is different from the depinning transitions found in
equilibrium systems with quenched disorder. Similar delocalization
transitions have recently been found in a variety of non-equilibrium
processes\cite{HLMP,HM,MKB,GM,MK}.

Let $P(h,t)$ denotes the fraction of agents with efficiency $h$ at time
$t$.  One can easily write down the evolution equation for $P(h,t)$
by counting all possible gain and loss terms. For $h\ge 1$, this
equation reads
\begin{eqnarray}
{dP(h,t)\over dt}=&-&P(h,t)
\sum_{h'=h+1}^{\infty}P(h',t)-(p+q)P(h,t)\nonumber\\
&+& q P(h+1,t)+pP(h-1,t)\nonumber \\
&+&P(h,t)\sum_{h'=0}^{h-1} P(h',t).
\label{pht}
\end{eqnarray}
In writing Eq.~(\ref{pht}), we have used the fact that when the total
number of agents diverges, the joint probability distribution
$P(h,h',t)$ of two agents having efficiencies $h$ and $h'$ factorises,
$P(h,h',t)=P(h,t)P(h',t)$, and the mean-field theory becomes exact.

It proves convenient to consider the cumulative distribution,
$F(h,t)=\sum_{h'\geq h} P(h',t)$.  {}From Eq.~(\ref{pht}), we
immediately derive the evolution equation for $F(h,t)$,
\begin{eqnarray}
{dF(h,t)\over dt}=&-&F^2(h,t)+(1-p-q)F(h,t) \nonumber\\
&+&q F(h+1,t)+pF(h-1,t).
\label{fht}
\end{eqnarray}  
Note that this equation is valid for all $h\ge 1$ and by the probability
sum rule we have $F(0,t)=1$ for arbitrary $t$. Also, $F(h,t)\to 0$ as
$h\to \infty$ for all $t$.

Equation (\ref{fht}) is a nonlinear difference-differential equation and
is, in general, hard to solve exactly.  Fortunately, many asymptotic
properties can be derived analytically without solving Eq.~(\ref{fht}).
First we note that $F(h,t)$ approaches a traveling wave form as it
follows e.g. from direct numerical integration of Eq.~(\ref{fht}).
Thus, we seek a solution of the form $F(h,t)=f(h-vt)$.  By inserting it
into Eq.~(\ref{fht}) we find that $f(x)$ satisfies
\begin{eqnarray}
-v {{df}\over {dx}}&=&-f^2(x)+(1-p-q)f(x)\nonumber\\
&+& qf(x+1)+pf(x-1),
\label{fx}
\end{eqnarray}
which should be solved subject to the boundary conditions $f(-\infty)=1$
and $f(\infty)=0$.  To determine $v$, we linearize Eq.~(\ref{fx}) in the
tail region, $x\to\infty$.  The resulting linear equation admits an
exponential solution, $f(x)\sim \exp(-\lambda x)$.  By inserting this
asymptotics into the linearized version of Eq.~(\ref{fx}) we find that
the growth rate $v(\lambda)$ is related to the decay exponent $\lambda$
via
\begin{equation}
v(\lambda)={1-p-q+qe^{-\lambda}+pe^{\lambda}\over \lambda}.
\label{vl}
\end{equation}  
Thus we have a family of eigenvalues parameterized by $\lambda$.
According to a general selection principle which applies to a wide class
of nonlinear equations\cite{Mur,Bram}, only one specific rate out of
this family of possible $v$'s is selected.  Usually, the minimum rate is
selected.  For sufficiently steep initial conditions, the minimum rate
is universal, while extended initial conditions might affect the
magnitude of the {\em admissible} minimum rate.

The function $v(\lambda)$ in Eq.~(\ref{vl}) has a unique
minimum at $\lambda=\lambda^{*}$ given by the solution of ${{dv}\over
  {d\lambda}}=0$, or
\begin{equation}
1-p-q+qe^{-\lambda^*}+pe^{\lambda^*}
=\lambda^*\left[-qe^{-\lambda^*}+pe^{\lambda^*}\right].
\label{lamb}
\end{equation}         
The corresponding minimum rate $v_{\rm min}(p,q)\equiv v(\lambda^{*})$
is given by Eq.~(\ref{vl}), or 
\begin{equation}
v_{\rm min}=-qe^{-\lambda^*}+pe^{\lambda^*}
\label{vel}
\end{equation}
as it follows from (\ref{lamb}).  An analysis of
Eqs.~(\ref{vl})--(\ref{vel}) shows that there exists a critical line
$p_c(q)$ in the $(p,q)$ plane,
\begin{equation}
\label{critic}
p_c(q)=\cases{1+q-2\sqrt{q} & for $q\ge 1$, \cr
              0  & for $q\le 1,$}
\end{equation}         
such that $v(\lambda)>0$ for all $\lambda\ge 0$ as long as $p>p_c(q)$.
For a fixed $q$, as $p\to p_c(q)$ from above, $v_{\rm min}\to 0$ and
for $0<p<p_c(q)$, the curve $v(\lambda)$ crosses zero at
$\lambda=\lambda_1$ and $\lambda=\lambda_2$ with $\lambda_2>\lambda_1$.
When $\lambda_1<\lambda<\lambda_2$, $v(\lambda)$ becomes negative. This
tells that there might be no traveling wave solution for $0<p<p_c(q)$ and we
anticipate that the efficiency distribution should become stationary.
Note that for $q<1$, $p_c(q)=0$ and this regime does not occur.

\begin{figure}
\narrowtext
\centerline{\epsfxsize\columnwidth\epsfbox{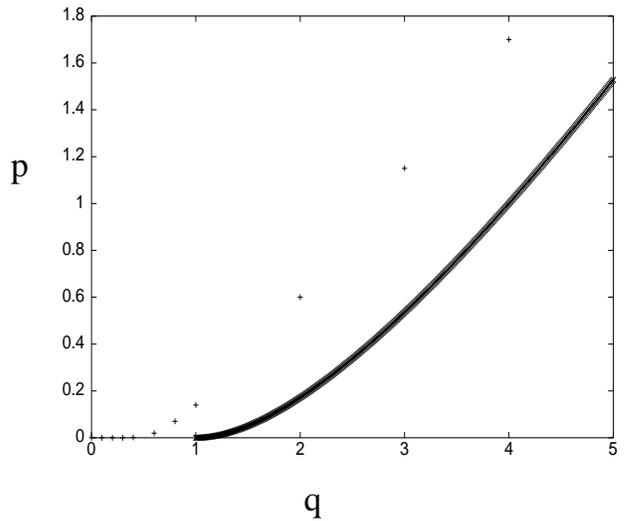}}
\caption{The thick line represents the critical locus, 
  $p_c(q)=1+q-2\sqrt{q}$, for the mean-field theory. The $+$'s indicate
  numerically obtained critical points in one dimension. For sharply
  decaying initial conditions, the economy is developing when $p>p_c(q)$
  and stagnant when $p\le p_c(q)$.}
\label{phd}
\end{figure}           

With the above picture in mind, we now discuss the selection
principle more carefully.  Consider an exponentially decaying initial
condition, $F(h,0)\sim e^{-\alpha h}$ with $\alpha>0$.  When $p>p_c(q)$,
the rate is positive for all $\lambda>0$ and $v(\lambda)$ has a
unique minimum at $\lambda=\lambda^*$.  Applying the selection
principle we find that for sufficiently steep initial conditions,
$\alpha>\lambda^*$, the selected growth rate is 
$v_{\rm min}=v(\lambda^*)$.  Consider now sufficiently extended initial
conditions, $\alpha<\lambda^*$.  In this case, $f(x)$ must decay at most
as $e^{-\alpha x}$ and therefore the growth rate is selected among
$v(\lambda)$, Eq.~(\ref{vl}), with $\lambda\leq \alpha$. The selection
principle now implies that the selected rate is $v=v(\alpha)$.

When $p\le p_c(q)$, we must separately consider two cases: $q> 1$ and
$q\le 1$.  For $q>1$, $v(\lambda)$ as given by Eq.~(\ref{vl}) becomes
negative in the region $\lambda_1<\lambda<\lambda_2$. We find that for
all $\alpha < \lambda_1$, the system still admits a traveling wave
solution and the selected rate is $v=v(\alpha)$.  However, for
$\alpha>\lambda_1$, the system no longer admits a traveling wave
solution. Instead, the distribution $F(h,t)$ reaches a stationary limit
$P_\infty(h)$ as $t\to \infty$.  By putting the time derivative equal to
zero on the left-hand side of Eq.~(\ref{fht}), we find that the
stationary efficiency distribution decays exponentially,
$F_\infty(h)\sim e^{-\mu h}$, with
\begin{equation}
\mu(p,q)=\ln\left[{-1+p+q+\sqrt{(1-p-q)^2-4pq}\over 2p}\right].
\label{mu}
\end{equation}
Note that $\mu(p,q)$ is real below the critical line, i.e., when $q>1$
and $p\leq p_c(q)$.  Interestingly, $\mu(p,q)$ remains finite on the
critical line $p=p_c(q)$. {}From Eqs.~(\ref{mu}) and (\ref{critic}) we
find that $\mu_c(q)=\mu(p_c(q),q)$ reads
\begin{equation}
\mu_c(q)=\ln\left({\sqrt{q}\over \sqrt{q}-1}\right).
\label{muc}
\end{equation}
For $q\le 1$, $p_c(q)=0$.  When $p\to 0$, we have $v_{\rm min}\to 0$ and
$\lambda^* \to \infty$.  The divergence of the decay exponent
$\lambda^*$ indicates that when $p=0$ and $q<1$, the system
still admits a traveling wave solution and the selected rate is
$v=v(\alpha)$ if we start with an exponentially decaying initial
condition, $F(h,0)\sim e^{-\alpha h}$.  Of course, for compact initial
conditions (i.e., when $F(h,0)=0$ for sufficiently large $h$), the
efficiency distribution becomes stationary in the long time limit.  We
have verified all the above assertions via direct numerical
integration of Eq.~(\ref{fht}).

Although one cannot provide {\em explicit} expressions for the growth
rate in the developing phase, near the critical line $p_c(q)$ the growth
rate considerably simplifies. First we note that on general scaling
grounds one would guess that above the critical line, the growth rate
$v_{\rm min}(p,q)$ should be a function of $p-p_c$ with critical
behavior
\begin{equation}
\label{guess}
v_{\rm min}\sim (p-p_c)^\beta.
\end{equation}         
The actual behavior is found by a straightforward analysis of
Eqs.~(\ref{vl})--(\ref{lamb}) to yield
\begin{equation}
\label{near}
v_{\rm min}\to 
\cases{{p-p_c(q)\over (\sqrt{q}-1)\mu_c(q)}  & for $q>1$,  \cr
\cr
       {4\sqrt{p}\over \ln(1/p)}             & for $q=1$,  \cr
\cr
       {1-q\over \ln(1/p)}                   & for $0\leq q<1$,\cr}
\end{equation}         
where $\mu_c(q)$ is given by Eq.~(\ref{mu}).  Equation (\ref{near})
implies that the mobility exponent $\beta$ in the scaling relation
(\ref{near}) is equal to $1,1/2$ and $0$ for $q>1,q=1$ and $q<1$,
respectively.  In the last situation ($q<1$ and $p\to 0$), the growth
rate still approaches to zero but it occurs in an extremely slow
inverse logarithmic fashion.

The relaxation of the growth rate $v(t)$ towards its asymptotic value
$v_{\rm min}$ exhibits an interesting algebraic behavior.  Specifically,
the leading correction is proportional to $t^{-1}$, the next is of order
$t^{-3/2}$, etc.  Similar $t^{-1}$ correction was first derived by
Bramson in the context of a reaction-diffusion equation\cite{Bram}, and
was subsequently re-derived and generalized by a number of
authors\cite{vS,B+D}. The next correction was recently derived by Ubert
and van Saarloos\cite{EvS}.  In contrast to Refs.\cite{Bram,vS,B+D,EvS},
we consider the difference-differential equation. Fortunately, the
techniques\cite{Bram,vS,B+D,EvS} still apply (compare \cite{MK,KM}), so
we do not detail the derivation.  Following for instance an approach of
Ref.\cite{EvS}, one finds
\begin{equation}
v(t)=v_{\rm min}-{3\over 2\lambda^*}\,t^{-1}+A\,t^{-3/2}
+{\cal O}(t^{-2}),
\label{vt}
\end{equation}
with
$A=3\pi^{1/2}\{2(qe^{-\lambda^*}+pe^{\lambda^*})\}^{-1/2}(\lambda^*)^{-2}$.
The explicitly displayed terms are {\em universal} -- they do not depend
on initial condition as long as it is steep enough [i.e., it falls off
faster than $e^{-\lambda^* x}$].  The following terms in Eq.~(\ref{vt})
starting from ${\cal O}(t^{-2})$ correction are non-universal.  Thus not
only any sufficiently steep initial profile relaxes to a unique profile,
the approach to that profile occurs along (asymptotically) unique
trajectory.  Note also that the very slow $t^{-1}$ relaxation of the
growth rate leads to a logarithmic correction to the average efficiency,
\begin{equation}
\langle h\rangle=v_{\rm min}t-{3\over 2\lambda^*}\,\ln t
+{\cal O}(1).
\label{avh}
\end{equation}

Thus, we have a delocalization transition across the critical line
$p_c(q)$ in the $(p,q)$ plane for sharply decaying initial conditions.
For such initial conditions, as long as $p>p_c(q)$, the economy is in
the developing phase with the average efficiency increasing as $\langle
h\rangle \approx v_{\rm min} t$, where the rate $v_{\rm min}$ given by
Eqs.~(\ref{vel}) and (\ref{lamb}). For $p\le p_c(q)$ with $q>1$, the
system is localized and $\langle h\rangle$ approaches a time-independent
constant in the long time limit. For $p=0$ and $q<1$, the economy is in
the developing phase for unbounded initial efficiency distributions,
with the rate of growth explicitly dependent on the initial condition.
For economically more relevant compact initial conditions, the regime
$p=0$ and $q<1$ belongs to the stagnant phase.  These results are
summarized in the phase diagram in Fig.~1.

The mean-field version of the efficiency model is natural in the
interconnected modern economy.  In economy with limited communication,
however, the efficiency model in a low dimensional space rather than in
the fully connected graph might be more appropriate. In this case,
agents are placed on a finite dimensional lattice. The microscopic
dynamical steps (i)--(iv) remain the same except that in move (i), the
agent $j$ is chosen to be one of the nearest neighbors of $i$. Unlike
the mean-field theory, the correlations between $h_i$'s at different
sites remain nonzero in finite dimensions even in the thermodynamic
limit. We have studied this model numerically in one dimension. The
results are shown for lattice size $L=1000$ (we verified that for such
large systems, the finite size effect is insignificant).  Once again,
there is a delocalization transition in the $(p,q)$ plane across a
critical line as shown in Fig.~1.  The efficiency distribution $P(h,t)$
at different times in both phases is presented in Fig.~2.

\begin{figure}
\narrowtext
\centerline{\epsfxsize\columnwidth\epsfbox{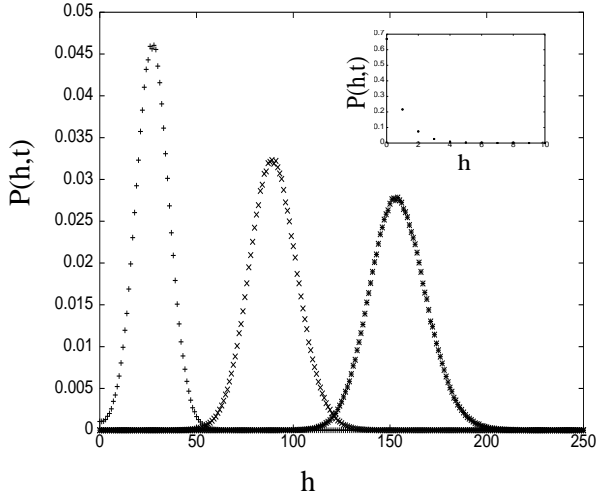}}
\caption{The distribution $P(h,t)$ at times $t=200$, $600$ and $1000$
in the moving phase for $p=3$, $q=4$. The inset shows the distribution
at the same times in the localized phase for $p=1$, $q=4$. For $q=4$, the
critical point is $p_c(q)\approx 1.7$.}
\label{3414}
\end{figure}           

We now stress important differences between mean-field and
finite-dimensional situations. In the former case the nature of the two
phases depends on the steepness parameter $\alpha$, while in one
dimension the nature of the final state is independent of $\alpha$. For
example, in the developing phase the system always has a traveling wave
solution with a rate that depends on $p$ and $q$ but does not depend
on $\alpha$.  We have tested this fact numerically for several values of
$\alpha$.  This result is rather counter intuitive as it suggests that
correlations seem to restore universality that mean-field theory
lacks.  Another important distinction is a very different behavior of
the width of the efficiency distribution in the developing phase.
Indeed, in mean-field the width is constant, while in one dimension it
increases with time [see Fig.~2].  Moreover, the width increases as a
power law, $w=\sqrt{\langle h^2\rangle -\langle h\rangle^2} \sim
t^{\beta}$ for large $t$, with $\beta\approx 0.31$ in $(1+1)$
dimensions.

In $(d+1)$-dimensions, one can interpret the efficiency $h_i(t)$ as the
height of a surface growing on a $d$-dimensional substrate. In this
language, our efficiency model represents a continuous time polynuclear
growth (PNG) model with additional adsorption and desorption
rates\cite{zhang}.  The continuous PNG model without desorption has been
studied within mean-field theory\cite{BBDK} and was found to be always
in the moving phase as expected.  {}From the general analogy to PNG
models, we expect that the moving phase in the efficiency model
corresponds to a growing interface belonging to the Kardar-Parisi-Zhang
(KPZ) universality class\cite{zhang}.  The numerically obtained width
exponent $\beta\approx 0.31$ in $(1+1)$-dimensions is consistent with
the KPZ prediction $\beta=1/3$. It would be interesting to determine the
universality class of the delocalization transition. Phase transitions
in several PNG models in $(1+1)$-dimensions belong to the directed
percolation (DP) universality class (see e.g.  Ref.\cite{KW}).  Other
similar growth models exhibit phase transitions that do not belong to
the DP universality class\cite{HLMP,MKB}. It remains an open question
whether the phase transition in the efficiency model in
$(1+1)$-dimensions belong to the DP universality class.

In summary, we have investigated a simple model of the dynamics of
efficiencies of competing agents.  The model takes into account
stochastic increase and decrease of the efficiency of every agent,
independent of other agents, and interaction between the agents which
equates the efficiencies of underachievers to that of better
performing agents. We have shown that the model displays a phase
transition from stagnant to growing economy.

\medskip
\noindent
One of us (PLK) acknowledges support from NSF (grant DMR9978902) and ARO
(grant DAAD19-99-1-0173).

\end{multicols}
\end{document}